\documentclass[prl,showpacs,twocolumn,amsmath,amssymb]{revtex4}
\usepackage[dvips]{graphicx}
\usepackage{epsfig}
\usepackage{psfrag}

\begin{document}
\title{Magnetic-field induced band-structure change in CeBiPt}
\author{N. Kozlova$^1$, J. Hagel$^2$, M. Doerr$^2$,
J. Wosnitza$^2$\cite{presadr}, D. Eckert$^1$, K.-H. M\"uller$^1$,
L. Schultz$^1$, I.~Opahle$^1$, S.~Elgazzar$^1$, Manuel Richter$^1$,
G. Goll$^3$, H. v. L\"ohneysen$^{3,4}$, G. Zwicknagl$^5$,
T. Yoshino$^6$, T. Takabatake$^6$}
\affiliation{
$^1$Institut f\"ur Festk\"orper- und Werkstoffforschung (IFW) Dresden,
D-01171 Dresden, Germany\\
$^2$Institut f\"ur Festk\"orperphysik, Technische Universit\"at
Dresden, D-01062 Dresden, Germany\\
$^3$Physikalisches Institut, Universit\"at Karlsruhe,
D-76128 Karlsruhe, Germany\\
$^4$Forschungszentrum Karlsruhe, Institut f\"ur Festk\"orperphysik,
D-76021 Karlsruhe, Germany\\
$^5$Institut f\"ur Mathematische Physik, Technische Universit\"at
Braunschweig, D-38106 Braunschweig, Germany\\
$^6$Department of Quantum Matter, ADSM, Hiroshima
University, Higashi-Hiroshima 739-8530, Japan}

\date{\today}

\begin{abstract}
We report on a field-induced change of the electronic band structure of
CeBiPt as evidenced by electrical-transport measurements in pulsed magnetic
fields. Above $\sim$25~T, the charge-carrier concentration increases
nearly 30\% with a concomitant disappearance of the Shubnikov--de Haas signal.
These features are intimately
related to the Ce 4$f$ electrons since for the non-$4f$ compound LaBiPt
the Fermi surface remains unaffected. Electronic band-structure
calculations point to a $4f$-polarization-induced change of the
Fermi-surface topology.
\end{abstract}


\maketitle
The influence of magnetic fields on the electronic band structure of
metals is usually minute and may, therefore, in most cases be
disregarded because of the different relevant energy
scales of the itinerant electrons in conventional metals and of the
applied magnetic fields: the typical Fermi energy is of the
order eV compared to a Zeeman energy of $\sim$6~meV at 50~T.
However, the situation may change considerably in case the relevant energy
scales of the electronic system are strongly reduced.

Prominent examples for field-induced changes of the Fermi surface
are, e.g., strongly correlated metals close to a ``quantum critical
point'' \cite{sac99}. Usually, the criticality results from
magnetic interactions leading to field-induced modifications of the
magnetic ground state \cite{bor04,kim03}. For some materials
different Fermi surfaces could be detected above and below
a metamagnetic phase transition \cite{jul94,bor04,aok93}. In the
paramagnetic state of a metal, however, a direct influence of an
external magnetic field on the electronic band structure is
usually not expected.

Here, we provide evidence for a rather sudden field-induced increase
of the charge-carrier concentration connected with a Fermi-surface
change in the semimetal CeBiPt. This remarkable phenomenon is absent
in the non-4$f$ sister compound LaBiPt. Both compounds belong to the
intermetallic series $R$BiPt ($R$ = rare-earth metal) that
shows a rich diversity of ground states depending on $R$
\cite{can91}. Explicitly, YbBiPt is a ``super''-heavy-fermion metal
\cite{fis91}, NdBiPt is a small-gap semiconductor \cite{can91},
LaBiPt is a superconductor with critical temperature $T_c =
0.9$~K, and CeBiPt is a commensurate antiferromagnet with
an ordering temperature of $T_N = 1.1$~K \cite{pie00,gol02}.

\begin{figure}
\centering
\includegraphics[width=8.3cm,clip=true]{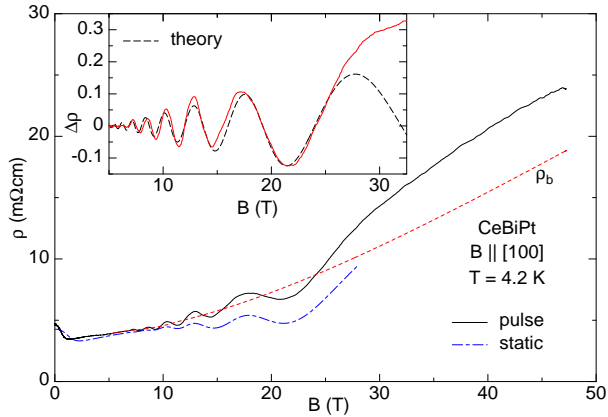}
\caption[]{Field dependence of the resistivity of CeBiPt at 4.2~K
in a pulsed field of up to 48~T and in a static field of up to 28~T.
The inset shows the SdH signal after subtracting the steady background
$\rho_b$ from the pulsed-field data. The dashed curve represents
the behavior expected theoretically.}
\label{rvsb}
\end{figure}

In the following we present high-field investigations of the
magnetization and of the
electrical-transport properties of LaBiPt and CeBiPt in the
paramagnetic metallic state. Both compounds are semimetals with very low
charge-carrier concentrations and correspondingly small Fermi surfaces
\cite{gol02}. The electronic properties of LaBiPt can consistently
be described by standard Fermi-liquid theory. The observed
magnetic quantum oscillations \cite{gol02,wos04} agree very
well with band-structure calculations \cite{ogu01}.
The picture changes considerably for the isostructural semimetal
CeBiPt. Above $\sim$10~K and below $\sim$20~T, the electronic properties
follow conventional Fermi-liquid theory as in LaBiPt. Towards
lower temperatures, however, the Fermi surface changes, i.e., for
certain field orientations the Shubnikov--de Haas (SdH) frequency
increases by almost a factor of two between 10 and 0.4~K \cite{gol02}.
Another unusual feature is found in pulsed-field experiments.
Above about 25~T the SdH signal vanishes and, instead, the
magnetoresistance increases
considerably (Fig.\ \ref{rvsb}) hinting at a field-induced
Fermi-surface modification \cite{wos04}. This feature is not related
to the quantum limit since at 25~T still about five Landau levels
(of both spin orientations) are occupied.

The single crystals of CeBiPt and LaBiPt were grown at Hiroshima
University by use of the Bridgman technique. Details of the crystal
growth have been reported elsewhere \cite{pie00,gol02}.
The electrical-transport and magnetization measurements were performed
at the High Magnetic Field Laboratory Dresden (HLD) in pulsed fields
up to about 50~T at temperatures above $T = 1.8$~K by use of a $^4$He
gas-flow cryostat. For the transport measurements six 40~$\mu$m gold
wires were attached to the samples with graphite paste. AC currents
up to 1~mA with frequencies between 10 and 50~kHz were applied for about
80~ms just before and during the field pulse. The reliability of the
data was checked for different currents and frequencies. Uncertainties
of the geometry factors result in error bars of about 20\% in the
absolute resistivity. The samples were tightly fixed to the sample
holder with IMI7031 varnish. Since small
misalignments of the contacts are unavoidable the longitudinal and
transverse (Hall) resistances were extracted from positive and negative
field pulses applied successively at each temperature \cite{rem1}.
The data shown in the following, therefore, comprise either the
completely symmetric (longitudinal) and antisymmetric (transverse)
signals.

The overall agreement between pulsed-field and static-field data is
very good (Fig.\ \ref{rvsb}) \cite{gol02}. The initial decrease of
the resistivity, $\rho$, reflects most probably
antiferromagnetic fluctuations above $T_N$ in the paramagnetic state
as evident from its absence at 20\,K \cite{wos04}. Then, up to about
25~T, the oscillations caused by the SdH effect appear. Towards higher
fields, however, instead of exhibiting further maxima and minima, $\rho$
just increases monotonically.

The inset in Fig\ \ref{rvsb} shows the expected SdH signal (dashed
curve) in comparison to the experimental data. For the theory curve
we used the well-known SdH formulas (see \cite{sho84} for details)
with parameters $F = 48$~T for the SdH frequency, $m_c = 0.24~m_e$ for
the effective mass, and $T_D = 2.7$~K for the Dingle temperature,
in good agreement with the static-field data
\cite{gol02}. For the experimental signal we plotted $\Delta\rho =
(\rho-\rho_b)/\rho_b$, with $\rho_b$ the steady background resistivity
shown by the dashed line in Fig.\ \ref{rvsb} \cite{rem4}. It is
evident that above 25~T the oscillating signal disappears. We should
note that there is some ambiguity determining
$\rho_b$. However, in order to recover an oscillating behavior
of $\Delta\rho$ above 25~T a highly artificial oscillatory background
would have to be assumed.

In previous experiments, it was shown that the resistance increase
can be followed up to 60~T and that the SdH oscillations vanish
independent of sample orientation in field \cite{wos04}. This
contrasts with the
temperature-dependent change of the Fermi surface that occurs only
for fields aligned within about 15~deg along the main cubic-lattice
axes \cite{gol02}.

\begin{figure}
\centering
\includegraphics[width=8.3cm,clip=true]{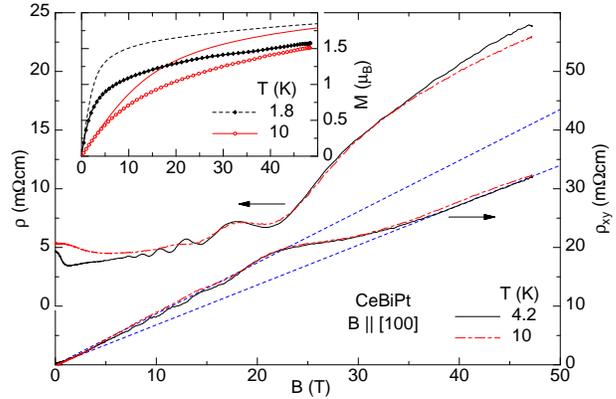}
\caption[]{Field dependence of the longitudinal and transverse resistivities
of CeBiPt for different temperatures. The dashed lines are linear fits
to the low and high-field Hall data. The inset shows the measured
(symbols) in comparison to the calculated magnetization (lines).}
\label{rhall}
\end{figure}

In order to investigate the high-field behavior of CeBiPt in more detail,
we measured the Hall effect. Indeed, as the most important result
of this study, Hall-effect measurements in pulsed fields reveal a
clear change of slope of the Hall signal at this field range
(Fig.\ \ref{rhall}). This effect was found to be temperature independent
between 1.8 and 10~K. Matching the field where the strong increase of
the longitudinal resistance sets in, the average Hall coefficient,
$R_H = \rho_{xy}/B$, decreases by about 28\% (difference in the slopes
of the two dashed lines in Fig.\ \ref{rhall}).
The fit to the low-field ($B \alt 22$~T) Hall data results in a
hole-like charge-carrier concentration of $n_h^{low} = (R_He)^{-1} =
7.2(3)\times 10^{17}$~cm$^{-3}$, whereas at high fields ($B \agt
38$~T) it increases to $n_h^{high} =
9.2(3)\times 10^{17}$~cm$^{-3}$. The low-field value
agrees well with the result of static-field measurements
($n_h^{low} = 7.7\times 10^{17}$~cm$^{-3}$ \cite{gol02}) in
view of the experimental error caused mainly by the
sample-geometry uncertainties.

It is this field-induced increase of the charge-carrier concentration
that is unique for the present paramagnetic metal. Earlier
band-structure calculations resulted in two small hole-like Fermi
surfaces at the Brillouin-zone center and even smaller
electron-like Fermi surfaces surrounding them \cite{gol02}.
Assuming a simple single-band picture, the low-field (below 25~T)
SdH results are in line with these calculations, although the smallness
of the electron pockets prohibits an experimental verification by our
SdH measurements. From theory it is not clear how this band structure
is modified above 25~T. A detailed analysis of the high-field
Fermi-surface topology from our data is excluded due to the lack of
any detectable SdH signal at these fields. The increasing number of
hole-like charge carriers would lead to a small increase (by about
18\%) of the SdH frequency.

\begin{figure}
\centering
\includegraphics[width=8.3cm,clip=true]{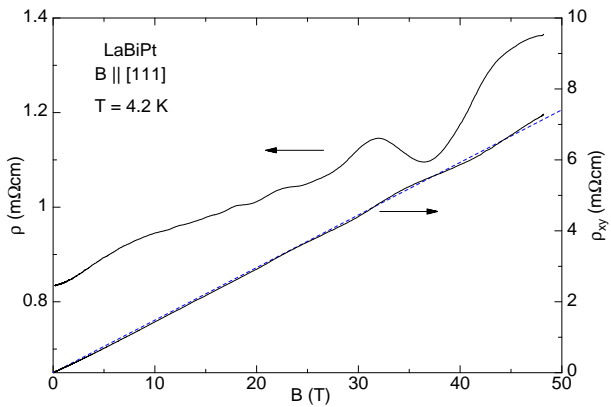}
\caption[]{Field dependence of the longitudinal and transverse resistivities
of LaBiPt. The dashed line is a linear fit to the Hall data.}
\label{rhlabipt}
\end{figure}

The temperature-dependent Fermi-surface change was found only for
CeBiPt, but not for the homologous non-4$f$ compound LaBiPt \cite{gol02}.
It was, therefore, straightforward to check for any unusual field-induced
phenomena in LaBiPt. As shown in Fig.\ \ref{rhlabipt}, for this metal
neither the longitudinal resistance nor the Hall effect reveal
any deviation from a well-behaved Fermi liquid. The SdH
signal for LaBiPt can be well explained \cite{wos04} by two slightly
different hole-like Fermi surfaces as predicted by band-structure
calculations \cite{ogu01}. Most notably, the transverse signal increases
-- except for small SdH traces -- linearly with magnetic field without any
unusual slope change. Consequently, the charge-carrier concentration in
LaBiPt remains at $n_h = 4.2(2)\times 10^{18}$~cm$^{-3}$ independent
of field.

These results clearly show that the Ce 4$f$ electrons are responsible
for the observed field-induced band-structure modifications.
A partial onset of $4f$ hybridization was suggested to explain the
temperature-dependent change of the Fermi-surface area observed
earlier \cite{gol02}. Thus, one possible scenario
for the observed feature would be that with
increasing magnetic field the $f$ electrons become decoupled from the
delocalized electrons. This may lead to a reduced number of electron-like
states which in turn would result in an increased value $n_h^{high}$ for
the hole-like charge-carrier concentration. A field-dependent decrease of
the $4f$ hybridization is well known from Ce-based heavy-fermion
compounds, such as CeCu$_6$ and CeB$_6$ \cite{fis92}. It should be noted,
however, that in contrast to these materials the Kondo effect appears to
play no significant role in CeBiPt. The Sommerfeld coefficient $\gamma$
deduced from specific-heat measurements is very low \cite{pie00}
and the effective mass from SdH measurements is only 0.24 free-electron
masses \cite{gol02}. We further observe a rather sudden change of the
charge-carrier concentration in CeBiPt that contrasts the smooth
effective-mass decrease in the above heavy-fermion systems.

As a consequence of the $f$ electrons loosing their itinerant character
one would expect a structure in the magnetization, $M$, with
increasing field, i.e., some increase of $M$ at about 25~T. We,
thus, measured $M$ of a large (374.5~mg) CeBiPt single crystal in
pulsed magnetic fields up to about 49~T. As shown in the
inset of Fig.\ \ref{rhall}, there is no significant feature
visible at 25~T in the magnetization $M(B)$. We therefore exclude
a field-induced decoupling of the $f$ electrons from the itinerant
charge carriers.

The expected magnetization curves at 1.8 and 10~K (lines in the
inset of Fig.\ \ref{rhall}) have been calculated assuming localized Ce
$4f$ moments. All necessary information for this
calculation is known; from inelastic neutron-scattering experiments
of CeBiPt powder a crystal-electric-field (CEF) splitting of 9.5~meV has
been measured \cite{sto04} and specific-heat data show that the quartet
$\Gamma_8$ state has the lowest energy \cite{pie00,lor04}. Since the
$\Gamma_7$ doublet is well separated from $\Gamma_8$ the free-ion
saturation magnetization of 2.14~$\mu_B$ cannot be reached up to 50~T.
The experimental data clearly fall below the expected $M$.

As an alternative scenario, the external field dependence of the band
structure was checked. Though the Fermi surfaces are tiny, the direct
Zeeman splitting of the band states is probably too small to explain
the effect. Field-induced $4f$ polarization, however, produces an
exchange field on the Ce $5d$ states that may yield band splittings
of $\sim$0.1~eV at 50~T in the saturated state.

At first, the zero-field band structure of CeBiPt was re-calculated
by use of a recent \cite{Opahle}, high-precision four-component
relativistic version of the full-potential local-orbital (FPLO)
code \cite{FPLO}. The Perdew--Zunger parametrization of the
exchange-correlation potential in the local spin-density approximation
(LSDA) was used. A basis set of optimized local orbitals with $5s$,
$5p$, $6s$, $6p$, and $5d$ states for Ce, Bi, and Pt
was used. The Ce $4f$ electron was not included in the valence basis
and localized by use of a confining potential.
A spherically averaged $4f$ charge and spin density was assumed.
Self-consistent calculations were performed with 8000~$k$-points in
the full Brillouin zone and 2000 Fourier components per atom for
the Ewald-potential representation.
The effect of all technical details of the calculation ($4f$ confinement,
spherical $4f$ charge density, other numerical parameters) has been
carefully checked to be below 10~meV in the region close to $E_F$.

\begin{figure}
\centering
\includegraphics[width=8cm,clip=true]{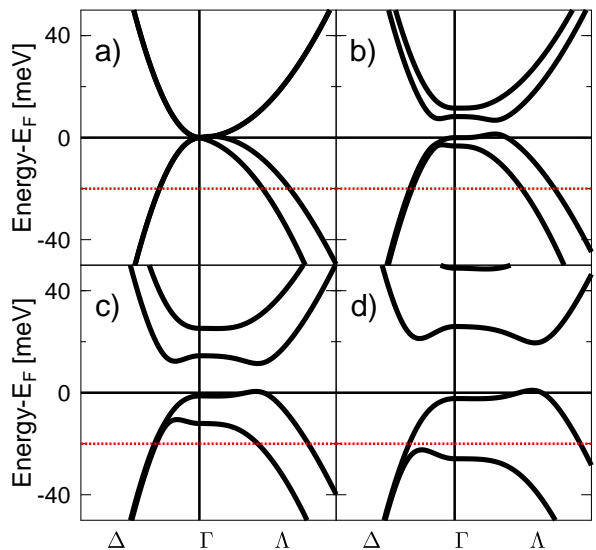}
\caption[]{LSDA bands of CeBiPt close to the Brillouin-zone center
along the symmetry lines $\Delta$ [from (0.1\,0\,0) to (0\,0\,0)
in units of $2\pi/a$] and $\Lambda$ [from (0\,0\,0) to
(0.067\,0.067\,0.067)]. The nominal $E_F$ is at zero energy,
the shifted Fermi level that yields the correct Fermi-surface area
is indicated at $-20$~meV by the dotted line.
(a) shows the non-magnetic case with unpolarized $4f$ shell;
In (b), (c), and (d), the $4f$ spin moment is fixed to be 0.2$\mu_B$,
0.5$\mu_B$, and 1.0$\mu_B$, respectively.}
\label{bands}
\end{figure}

The resulting band structure [Fig.\ \ref{bands}(a)] is very similar
to that published earlier \cite{gol02}, but the semi-metallic
character is yet more prominent with almost vanishing Fermi surfaces.
The experimentally observed hole-pocket areas can be matched if the
Fermi level is shifted by 20~meV to lower energy. This shift corresponds
to a tiny change of the valence electron number, $\Delta N/N \approx
10^{-5}$.
Such an uncertainty can be related either to a minute off-stoichiometry
of the samples below the limit of detectability or to the limited
accuracy of the LSDA used to calculate the single-electron levels.

Next, we note that all states in the vicinity of the Fermi level
($\pm 0.5$~eV) have predominantly Ce-$5d$ character. If $4f$ polarization
is applied (using the so-called open-core approach), intra-atomic exchange
interaction splits the Ce-$5d$ bands close to the center of the Brillouin
zone [Fig. \ref{bands}(b-d)]. The nominal Fermi level (integer number of
valence electrons) is taken as reference (0~eV) and a shifted Fermi
level that provides the correct Fermi-surface area is indicated
by the dotted line. The $4f$ spin polarization is gradually enhanced from
compensated spin densities [Fig.\ \ref{bands}(a)] to completely polarized
$4f$ state [Fig.\ \ref{bands}(d)]. The latter situation corresponds to a
field-saturated paramagnet.
With increasing $4f$ polarization, the electronic bands close to the
Brillouin-zone center are split in a Zeeman-like fashion into four
equidistant levels. Their splitting increases linearly with the
$4f$ spin moment up to $\sim$75~meV. The lowest of the four levels
crosses the shifted Fermi level before the $4f$ moment is saturated.
The related change of the Fermi-surface topology could explain
the measured Hall-effect peculiarity.
It should be noted that this scenario, though being attractive by its
simplicity, does not explain the observed temperature independence
of the effect.

Conduction-electron scattering off virtual CEF excitations provides an
alternative qualitative explanation. These processes enhance the effective
masses and shift the conduction-electron energies. The shifts, being
negligible in conventional metals, can affect the Fermi surfaces in
semimetals with low carrier concentrations. The resulting field-induced
changes should vary rather weakly with temperature as will be discussed
elsewhere.

In conclusion, we have presented evidence for a drastic change of the
electronic band structure of CeBiPt at $\sim$25~T. Above this
field the SdH oscillations vanish and the hole-like charge-carrier
concentration increases by about 28\%. The absence of these features
in LaBiPt clearly reflects the relevance of the $4f$ electrons for the
observed behavior. A splitting of the Ce-$5d$ bands close to $E_F$
due to exchange interaction with the polarized $4f$ states is a possible
explanation. This mechanism would yield a magnetic-field driven
metal-insulator transition for exact stoichiometry.

We would like to thank O. Stockert, M. Rotter, and A. M\"obius for
inspiring discussions. The work at Dresden was supported by the DFG
through SFB\,463 and the BMBF (FKZ 035C5 DRE). The work at Hiroshima was
supported by a grant for the International Joint Research Project NEDO and
the COE Research (13E2002) in a Grant-in-Aid from the Ministry of Education,
Culture, Sports, Science, and Technology of Japan. The work at Karlsruhe
was supported by the DFG through SFB\,195.


\begin{thebibliography}{123}
\bibitem{rem1}
For CeBiPt, due to technical reasons, the maximum negative field was
limited to -17~T. Anyway, a reliable disentanglement of the longitudinal
and transverse transport contributions was still possible.
\end{thebibliography}
\end{document}